\documentclass[final,3p,times,twocolumn]{elsarticle}

\usepackage{amssymb}

\journal{Chaos, Solitons and Fractals}

\begin{document}

\begin{frontmatter}



\title{Scaling Laws and Convergence Dynamics in a Dissipative Kicked Rotator}
\author{Danilo S. Rando$^{\rm 1,2}$}
\author{Edson D. Leonel$^{\rm 1}$}
\ead{edson-denis.leonel@unesp.br}
\author{Diego F. M. Oliveira$^{\rm 2}$}
\ead{diegofregolente@gmail.com}

\address{$^{\rm 1}$Departamento de F\'isica -- Instituto de Geoci\^encias e
Ci\^encias Exatas -- Universidade Estadual Paulista\\
Av.24A, 1515 -- Bela Vista -- CEP: 13506-700 -- Rio Claro -- SP --
Brazil\\
$^{\rm 2}$School of Electrical Engineering and Computer Science - University of North Dakota, Grand Forks, ND, USA.}

\begin{abstract}
The kicked rotator model is an essential paradigm in nonlinear dynamics, helping us understand the emergence of chaos and bifurcations in dynamical systems. In this study, we analyze a two-dimensional kicked rotator model considering  a homogeneous and generalized function approach to describe the convergence dynamics towards a stationary state. By examining the behavior of critical exponents and scaling laws, we demonstrate the universal nature of convergence dynamics. Specifically, we highlight the significance of the period-doubling bifurcation, showing that the critical exponents governing the convergence dynamics are consistent with those seen in other models.
\end{abstract}

\begin{keyword}
Chaos \sep Attractors \sep Kicked Rotator \sep Lyapunov



\end{keyword}

\end{frontmatter}




\section{Introduction}
Bifurcations are fundamental in the field of nonlinear dynamics \cite{hirsch2013differential}, providing crucial insights into the qualitative changes in system behavior as parameters vary \cite{alligood1998chaos}. These transformations, marked by the emergence of new stable states, periodic orbits, or chaotic regimes, are essential for understanding the complex dynamics of nonlinear systems across various scientific fields \cite{hogan2002nonlinear,orel2019non,li2022adaptive}. In this study, we examine bifurcations within the context of nonlinear dynamics \cite{alligood1998chaos,oliveira2008feigenbaum}, focusing on their theoretical foundations. Bifurcations are critical points where the qualitative behavior of a dynamical system undergoes significant changes \cite{simpson2009simultaneous}. These changes, caused by tiny variations in system's  parameters, lead to different dynamical regimes characterized by stability, oscillations, or chaotic behavior \cite{lai2011transient}. Bifurcations are vital indicators of system behavior, providing valuable insights into the mechanisms driving the complexity of nonlinear dynamics.
 
In nonlinear dynamics, bifurcations are described through mathematical expressions, revealing the qualitative changes in system behavior as parameters vary \cite{kim2011advanced}. Various classification schemes categorize bifurcations based on the type of qualitative change they induce. Common types include saddle-node bifurcations, pitchfork bifurcations, Hopf bifurcations, and period-doubling bifurcations, each characterized by distinct mathematical properties and dynamical behaviors \cite{strogatz2018nonlinear,guckenheimer2013nonlinear,kuznetsov1998elements}. Near bifurcation points, complex dynamics arise from the interaction between stable and unstable manifolds \cite{govaerts2000numerical}. Small perturbations in system parameters can lead to the emergence of new stable states or initiate chaotic dynamics \cite{strogatz2018nonlinear}. Understanding the behavior near bifurcation points is crucial for predicting and analyzing system behavior and for formulating effective control strategies to stabilize desired states during dynamical transitions \cite{govaerts2000numerical}.

The study of bifurcations in nonlinear dynamics has significant implications across various scientific disciplines, including physics \cite{vsuminas2017multi}, biology \cite{may1976simple}, and engineering \cite{han2023nonlinear}. In physics, bifurcation analysis helps reveal the behavior of complex systems such as fluid dynamics \cite{holmes2012turbulence}, plasma physics \cite{escande2016contributions}, and celestial mechanics \cite{valli2013nonlinear}. In biology, bifurcation theory provides insights into the dynamics of biological systems, including neuronal networks \cite{ashwin2024network}, ecological systems \cite{caravaggio2018nonlinear}, and genetic regulatory networks \cite{rosenfeld2009characteristics}. Additionally, bifurcation analysis has practical applications in engineering and control systems \cite{chen2012bifurcation}, aiding in the design of strategies to stabilize desired states or induce specific dynamical behaviors.

In this work, we revisit the problem of a dissipative kicked rotator model \cite{oliveira2014statistical} described by a two-dimensional nonlinear mapping. Similar to other systems, such as the renowned Fermi-Ulam model \cite{s2023bifurcations,ulam1961some,oliveira2013some}, an orbit diagram reveals transitions toward chaos via period-doubling bifurcations. Our focus is on analyzing convergence precisely at the bifurcation point, which is governed by a homogeneous and generalized function. The emergence of a scaling law corresponds to the synchronization of this homogeneous function with a predefined set of scaling hypotheses \cite{chen2020scaling,oliveira2018scaling}.

Within this analytical framework, the convergence dynamics of the stationary state reveal a delicate interplay of three critical exponents, intricately intertwined through a scaling law \cite{teixeira2015convergence}. However, as the system approaches the bifurcation point, its dynamics deviate from the homogeneous function, leading to the emergence of exponential decay patterns, as discussed in previous studies \cite{teixeira2015convergence}. The temporal decay relaxation exhibits a clear power-law dependence, influenced by the proximity in parameter space to the bifurcation point.

The paper is organized as follows: Section II provides an overview of the system and introduces the equations governing its dynamics. Section III explores the evolution of the stationary state near a period-doubling bifurcation. This examination begins with characterizing the bifurcation and then delves into analyzing its nearby evolution. Concluding remarks are presented in Section IV.

\section{The model and the mapping}
\label{sec2}

The kicked rotator model \cite{levi2003quantum,berry1984incommensurability} serves as a fundamental cornerstone in the field of nonlinear dynamics, providing a simplified yet insightful representation of complex dynamical systems \cite{kaplan1998linear}. At its core, the kicked rotator model captures the interplay between deterministic dynamics and periodic forcing, which is the predictable behavior of the system based on its initial conditions. This interaction is crucial because it can lead to complex and chaotic behaviors, even in a system that, in principle, could be predictable. This emergent complexity makes the kicked rotator model a classic example in the study of chaos and dynamical systems theory, resulting in rich and diverse dynamical behaviors \cite{balazs1986chaos}.

Mathematically, the kicked rotator model can be described by a two-dimensional mapping  \cite{bogomolny1988smoothed,bohigas1993manifestations} for the variables $(\theta,I)$ evaluated from the instant $n$ to the instant $n+1$. The dissipative relativistic standard map can be written as
\begin{equation}
S:\left\{\begin{array}{ll}
	\theta_{n+1}=[\theta_n + \frac{I_n}{\sqrt{1+(\rho I_n)^2}}-\xi I_n] ~~ {\rm mod~2\pi}~~\\
	I_{n+1}=(1-\psi)I_n + Ksin(\theta_{n+1})	
\end{array}
\right.~,
\label{eq1}
\end{equation}
where the parameter $K$ controls the transition from integrable ($K=0$) to non-integrable ($K \neq 0$). The parameter $\rho$ controls the shift from Newtonian to relativistic dynamics. When $\rho \rightarrow 0$, it reduces to the Newtonian standard map. On the other hand, in the ultra-relativistic limit ($\rho \rightarrow \infty$), the system tends to be integrable, as shown in the map (\ref{eq1}) for non-dissipative dynamics \cite{perre2020dynamics}.

We also introduce a dissipative parameter, $\psi$, which is in the range $\psi \in [0,1]$. This parameter shows the fraction of energy lost by particles at each step. It is important to note that when both $\psi$ and another parameter, $\xi$, are zero, all results related to the Hamiltonian area-preserving relativistic standard map are restored.

\begin{figure}[t]
	\centerline{\includegraphics[width=1.0\linewidth]{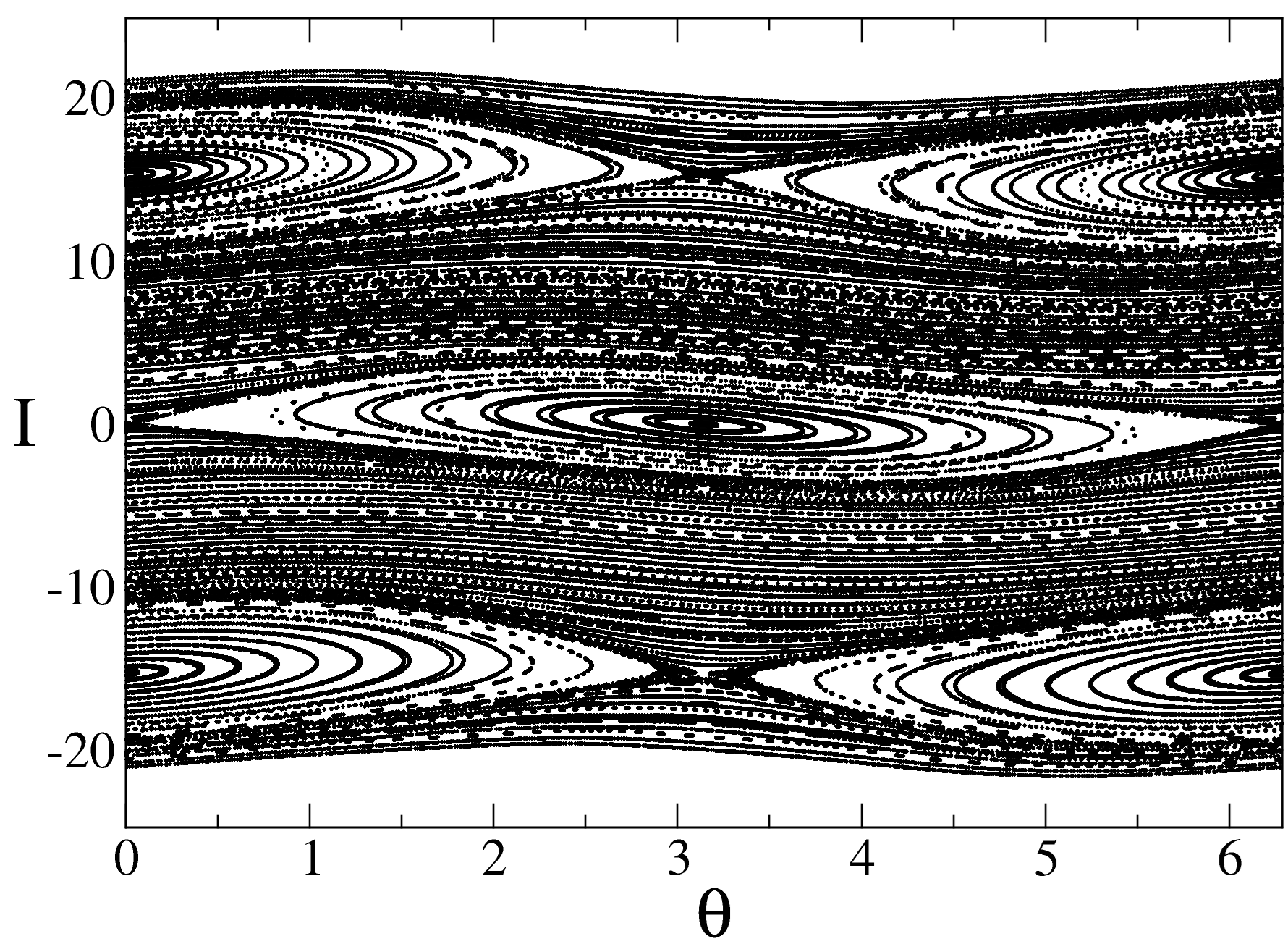}}
	\caption{Plot of the phase space considering the parameters $\psi = 0$, $\xi = 0.4$, $\rho = 0.15$ and $K=1.55$.} 
	\label{Fig1}
\end{figure}

First, let us consider the conservative case where  $\psi=0$ and $\xi \ne 0 $. In this case, the system exhibits a mixed phase space featuring KAM (Kolmogorov-Arnold-Moser) islands, chaotic regions, and invariant spanning curves, as illustrated in Figure  \ref{Fig1}.
The conservative case in dynamical systems means that energy is preserved over time, so the volume in phase space does not change. In these systems, trajectories can be quasi-periodic, regularly returning to similar regions of phase space, or chaotic, but always confined to specific regions. Analyzing these systems is important for understanding the stability and predictability of their long-term behaviors.

\begin{figure}
    \centering
    \includegraphics[width=1.0\linewidth]{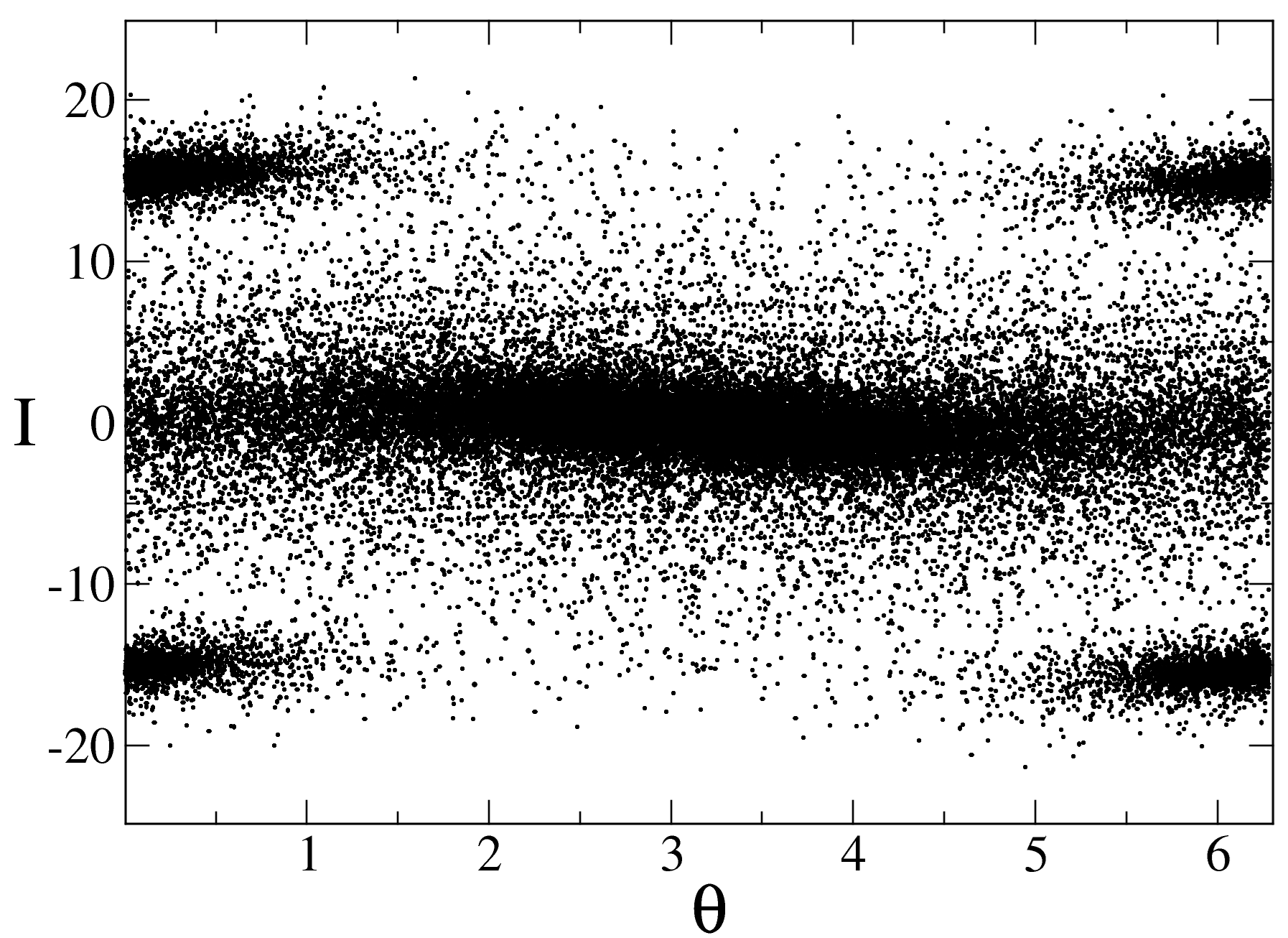}
    \caption{Plot of few orbits showing convergence to stationary dynamics considering $\psi = 0.01$, $\xi = 0.4$, $\rho = 0.15$ and $K=1.55$}
    \label{dissipative}
\end{figure}

The importance of introducing dissipation in a two-dimensional mapping, as in dynamical systems in general, lies in understanding the transition from regular to chaotic behavior. In a two-dimensional system without dissipation, the dynamics is conservative, meaning the volume in phase space is preserved over time. Typical trajectories are quasi-periodic or chaotic but remain within well-defined limits. However, when dissipation is included, this dynamics changes fundamentally, leading to a contraction in phase space and the emergence of attractors, as shown in Fig. \ref{dissipative}. This not only changes the behavior of the system but can also introduce complex behaviors. Studying these changes is crucial to understand how real systems, which are naturally dissipative, evolve over time and what patterns of behavior they may show.

\begin{figure}[h]
	\centerline{\includegraphics[width=0.75\linewidth]{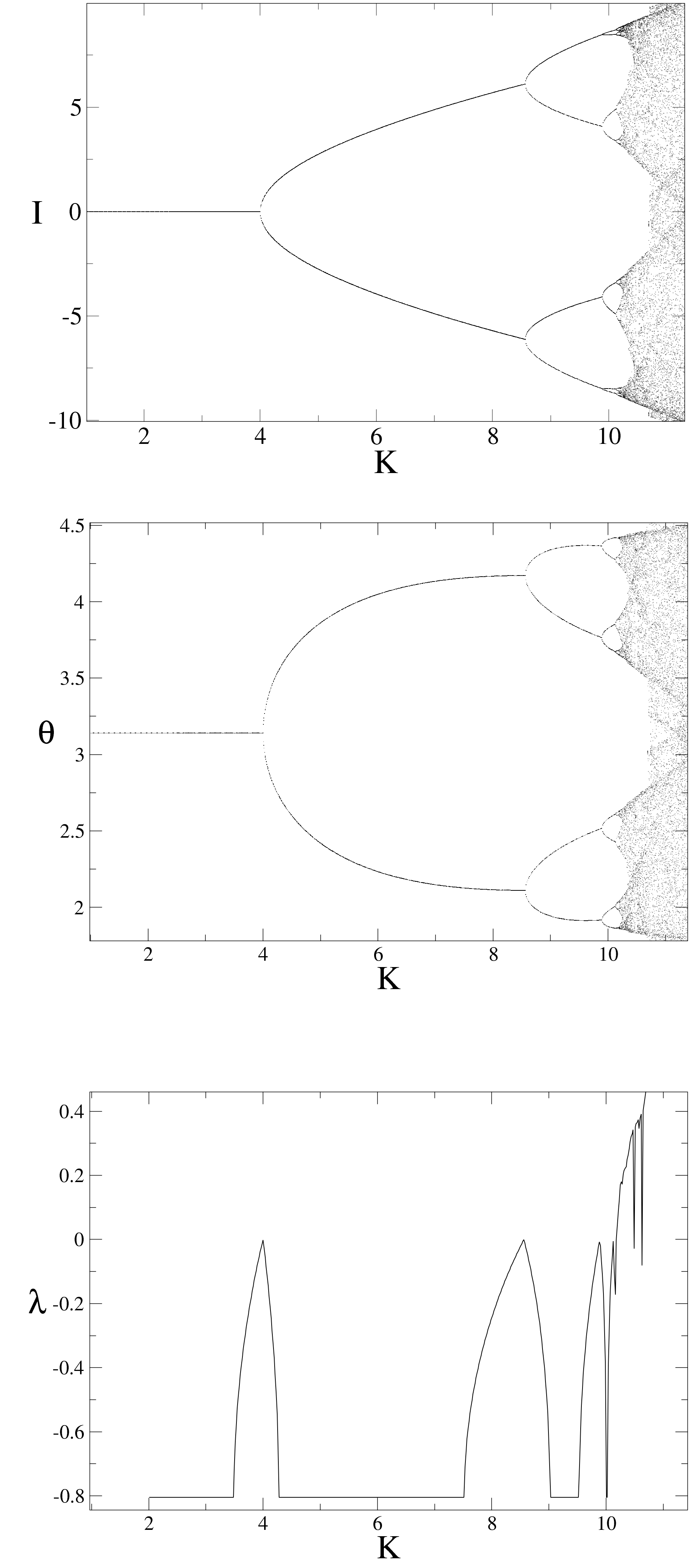}}
	\caption{The bifurcation cascade is displayed for (a) $I$ and (b) $\theta$, both plotted against the parameter $K$. In panel (c), the Lyapunov exponent associated with (a) and (b) is shown. In this analysis, we have fixed the parameters $\psi = 0.01$, $\xi = 0.4$, and $\rho = 0.15$. }
	\label{Fig2}
\end{figure}

Once dissipation is added to the model, our goal is to investigate the dynamics around the bifurcation points. These bifurcations can create new structures in phase space, change attractors, and generally transform the dynamic behavior of the system. In our case, the changes happens as the one changes the parameter $K$.
Figure \ref{Fig2} (a-b) shows the bifurcation diagrams for the variables  $I$, $\theta$ from the mapping (\ref{eq1}), considering the dissipative case with $\psi = 0.01$, $\xi = 0.4$, $\rho = 0.15$, as a function of the parameter $K$. On the other hand, Figure \ref{Fig2} (c) shows the behaviour of the Lyapunov exponent for the same range of K as seen in Figure \ref{Fig2} (a-b). The Lyapunov exponent is an important tool for understanding the stability and chaotic behavior of dynamical systems. 
 As discussed by Eckmann and Ruelle \cite{eckmann1985ergodic}, the Lyapunov exponents are defined as:
\begin{equation}
\lambda_j=\lim_{n\rightarrow\infty}{1\over{n}}\ln|\Lambda_j|~~,~~j=1,
2~~,
\label{eq4}
\end{equation}
where $\Lambda_j$ are the eigenvalues of
$M=\prod_{i=1}^nJ_i(x_i,y_i)$ and $J_i$ is the Jacobian matrix
evaluated over the orbit $(x_i,y_i)$. However, a direct
implementation of a computational algorithm to evaluate Eq. (\ref{eq4})
has a severe limitation to obtain $M$. For the limit of short $n$,
the components of $M$ can assume different orders of magnitude for
chaotic orbits and periodic attractors, making the implementation of
the algorithm impracticable. To avoid such a problem, $J$ can be
written as $J=\Theta T$ where $\Theta$ is an orthogonal matrix and $T$
is a right-up triangular matrix. $M$ is rewritten as $M=J_nJ_{n-1}\ldots
J_2\Theta_1\Theta_1^{-1}J_1$, where $T_1=\Theta_1^{-1}J_1$. A product of
$J_2\Theta_1$ defines a new $J_2^{\prime}$. In the next step, one can
show that $M=J_nJ_{n-1}\ldots J_3\Theta_2\Theta_2^{-1}J_2^{\prime}T_1$.
The same procedure can be used to obtain $T_2=\Theta_2^{-1}J_2^{\prime}$
and so on. Using this procedure, the problem is reduced to evaluate the
diagonal elements of $T_i:T_{11}^i,T_{22}^i$. Finally, the Lyapunov
exponents are given by
\begin{equation}
\lambda_j=\lim_{n\rightarrow\infty}{1\over{n}}\sum_{i=1}^n
\ln|T_{jj}^i|~~,~~j=1,2~~.
\label{eq005}
\end{equation}
If at least one of the $\lambda_j$  is positive, the orbit is said to be chaotic, and a it is zero at when a bifurcation happen. On the other hand, when the value of $\lambda_j$ is zero, it indicates that a bifurcation is happening in the system.

\section{Convergence to the stationary estate}
\label{sec3}
The convergence to a stationary state is important for understanding the behavior and stability of dynamical systems \cite{hirsch1988stability,lasalle1976stability}. It refers to the way that a system's state variables tend to stabilize around specific values or trajectories over time. This provides insights into the long-term behavior and stability properties of nonlinear systems \cite{hirsch2006monotone,drazin1992nonlinear}.

The process of convergence to a stationary state can happen in different ways, depending on the characteristics and dynamics of the system \cite{stuart1997convergence}. It may involve a gradual decay of transient behavior until the system reaches a stable equilibrium or periodic orbit. Alternatively, it can lead to the emergence of complex, non-periodic behavior, like chaotic attractors \cite{matsumoto1984chaotic,liu2004new}, where the state variables of the system show intricate and unpredictable patterns over time.

In our analysis of the stationary state for a period-doubling bifurcation, we focus on the $(I ~vs.~\theta)$ plane, where the dynamics develop. To quantify how close the particle is to the stationary state, it is essential to establish a metric that includes the particle's distance from the fixed point coordinates in our measurement. Therefore, we define:
\begin{equation}
d(n)=\sqrt{(I_n-I^*)^2+(\theta_n-\theta^*)^2}.
\label{eq_d}
\end{equation}

This definition allows us to measure the distance between system trajectories and the stationary state, which helps us understand the convergence dynamics toward the bifurcation point.

\subsection{Convergence at the bifurcation}

When we consider the convergence to the stationary state during bifurcations, we are interested in understanding how the system behaves as it approaches and stabilizes at stationary states near these critical points in the parameter space. This phenomenon is particularly interesting because it helps us understand how the system's behavior changes in response to small variations in its parameters, which can lead to transitions between different qualitative states. Understanding this convergence behavior is essential for predicting and analyzing the behavior of dynamical systems, especially in complex systems where bifurcations have a significant role.

To explain the convergence to stationary states during bifurcations, we can use homogeneous and generalized functions, which provide strong mathematical tools for capturing and analyzing the intrinsic dynamics. This is shown by the equation:

\begin{equation}
d(n,d_0)=\ell d(\ell^a n,\ell^b d_0),
\label{homog}
\end{equation}
where $ell$ is a scaling factor, $a$ and $b$ are characteristic exponents.

\begin{figure}[t]
\centerline{\includegraphics[width=1.0\linewidth]{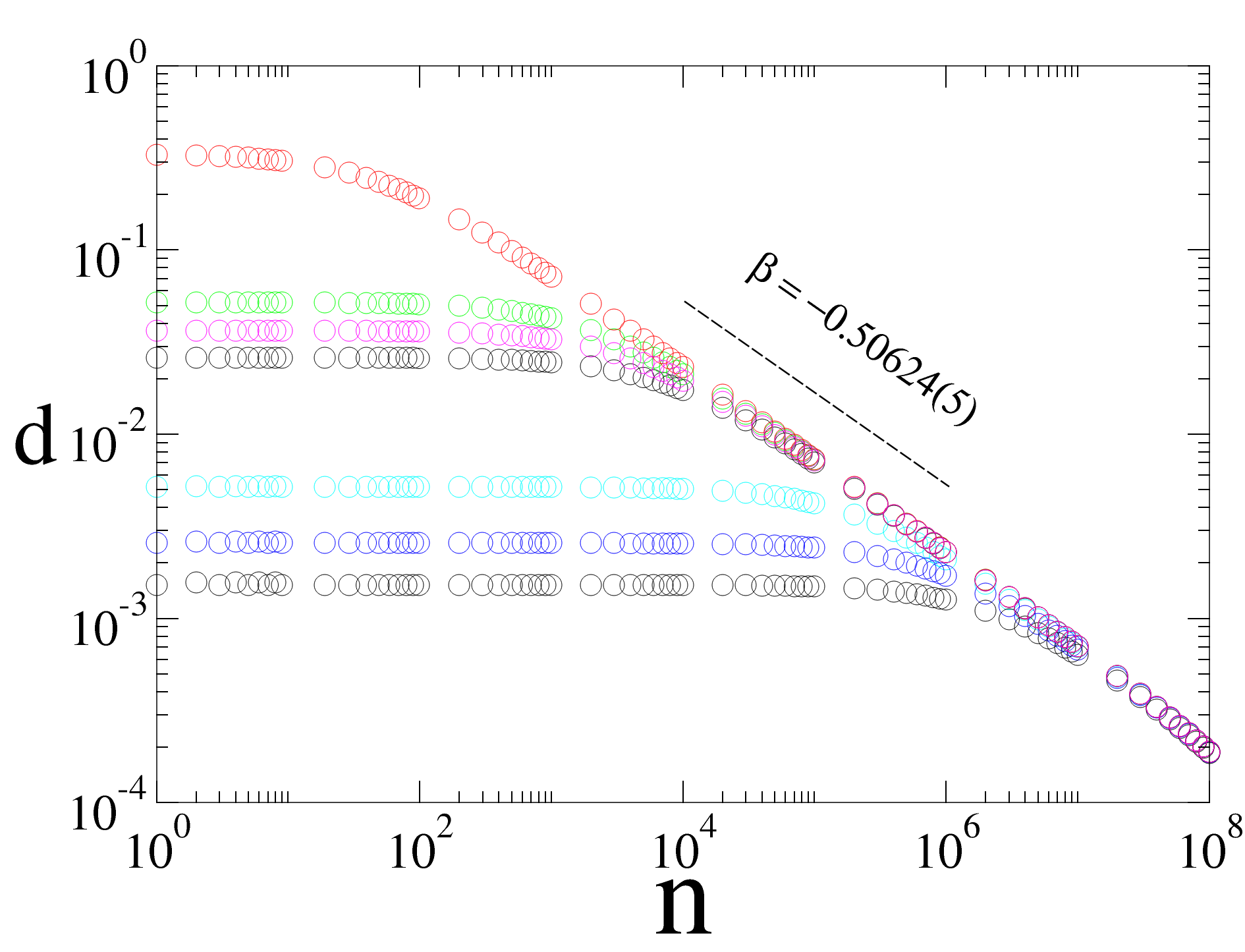}}
\caption{Behaviour of $d~vs.~n$ for different initial distances from the bifurcation point, the initial  conditions are associated with the distance $d$. } 
\label{Fig5}
\end{figure}

The lines plotted in Figure \ref{Fig5} show different behaviors, each with unique features. Initially, when $n$ is smaller than the crossover point $n_x$, there is a constant plateau. This plateau, typical of how the system behaves, follows the pattern $d(n) \propto d_0^{\alpha}$, where $\alpha$ is the exponent for the constant plateau. Interestingly, the constant plateau around $d_0$ is observed for several orders of magnitude, strongly suggesting that the exponent for the constant plateau $\alpha$ is equal to 1. After this, we see a change in behavior from a plateau to a power-law decay, which is characterized by the crossover point $n_x$. This decaying regime can be described by $d(n) \propto n^{\beta}$, where $\beta$ is the decaying exponent. After fitting a power law to the decaying region in Figure \ref{Fig5}, we find that $\beta = -0.50624(5) \approx -1/2$. Finally, the crossover number $n_x$, marking the transition from the constant plateau to the power-law decay, can be determined by assuming that at the crossover point, we have $n_x \propto d_0^z$, where $z$ is the crossover exponent.

 Observe that the equation (\ref{homog}) captures an important aspect of the behavior of the system. It describes a homogeneous relationship, showing how the quantity \( d(n,d_0) \) is related to its scaled versions \( d(\ell^a n,\ell^b d_0) \). In this context, \( \ell \) is a key scaling factor that influences the scaling dynamics, while \( a \) and \( b \) act as characteristic exponents that determine how the system responds to parameter changes. These exponents provide essential information about the system's sensitivity to variations in its parameters, giving insights into its underlying dynamics. By examining two different scaling hypotheses, namely \( \ell^a n = 1 \) and \( \ell^b d_0 = 1 \), we can find the relationships between \( \ell \), \( n \), and \( d_0 \), leading to expressions for \( \alpha \) and \( \beta \). These formulations come from comparing the scaling hypotheses with well-known scaling laws that govern various aspects of the system's behavior. The resulting scaling law, \( n_x = d_0^z \), where \( z = \frac{\alpha}{\beta} \), not only summarizes how the system behaves near the crossover point but also acts as a powerful tool for predicting its behavior under different conditions.
In our case, where \( \alpha = 1 \) and the critical exponent \( \beta = -1/2 \), this leads us to find the critical or crossover exponent.
Our main aim is to study the process of convergence to the stationary state during a bifurcation event. We specifically focus on understanding the complexities of a period-doubling bifurcation, which happens at the critical value of \( K_c = 3.999999997 \).

\subsection{Convergence near the bifurcation}

When we look at the convergence around the bifurcation point, the dynamics change from a uniform and generalized function to a behavior that looks like exponential decay, as seen in \cite{valli2013nonlinear}. In this case, the distance can now be described by the equation
\begin{equation}
d(n) = d_0 e^{-n/\tau},
\label{d_expo}
\end{equation}
The decay pattern in Equation (\ref{d_expo}) governs how the system moves towards the stationary state. Here, \( d_0 \) is the initial distance from the bifurcation point, and \( \tau \) indicates the relaxation time. It is important to note that \( \tau \) is proportional to \( \mu^{\delta} \), where \( \mu = K_c - K \) for \( K < K_c \). To find the relaxation time \( \tau \), we need to do extensive numerical simulations. 

We start by setting initial conditions at the boundary of the basin of attraction of the fixed point and let the dynamics evolve. When the distance from the fixed point reaches a certain tolerance, usually \( 10^{-6} \), we stop the simulation and record the number of iterations. This process is repeated for a large number of \( 10^6 \) different initial conditions. By averaging the relaxation times from this large set, we can determine the average relaxation time. We repeat this procedure for different values of the control parameters. The resulting data gives insights into the dynamics near the bifurcation point, which helps us to understand better the convergence behavior.

\begin{figure}[t]
\centerline{\includegraphics[width=1.0\linewidth]{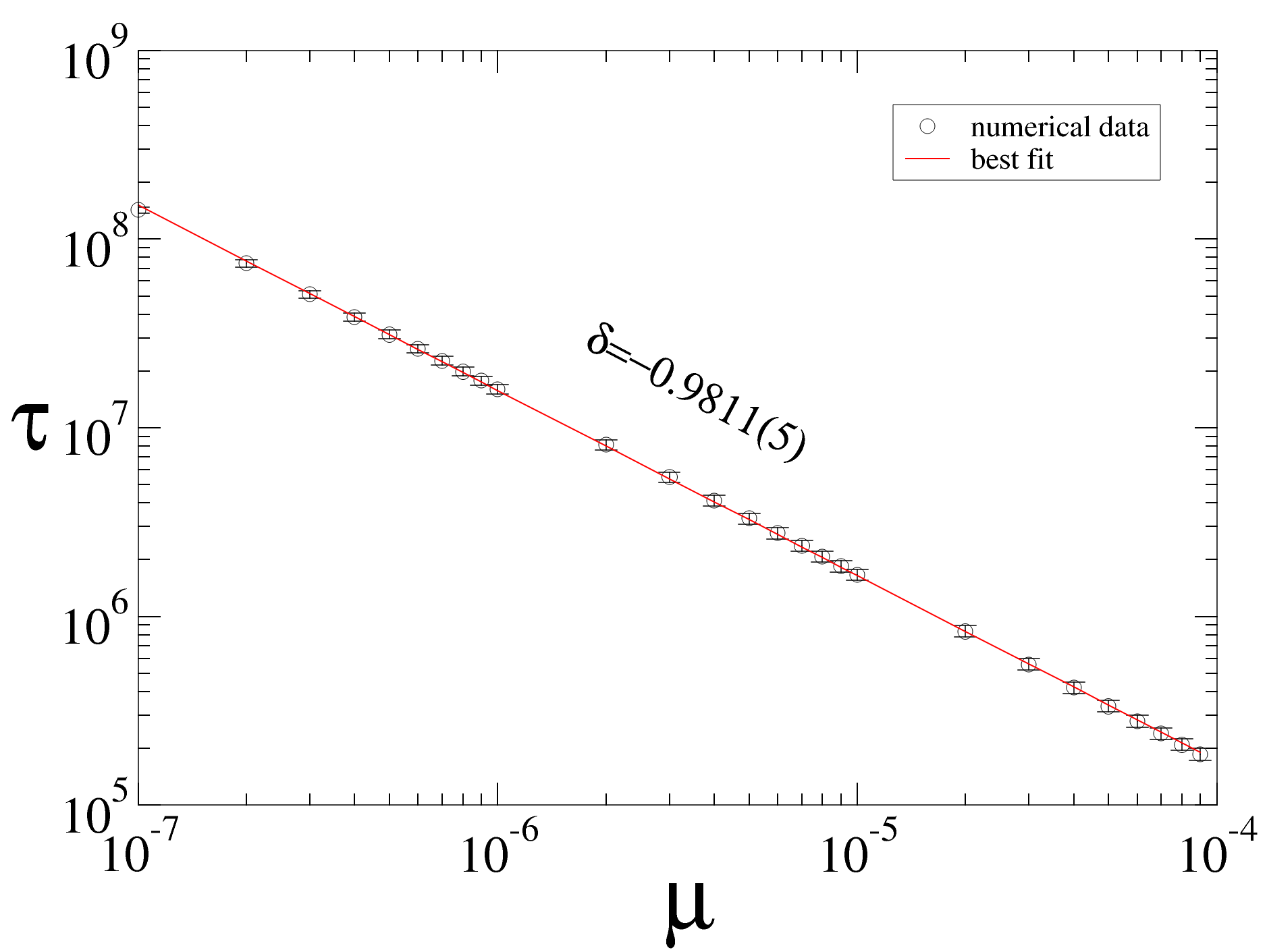}}
\caption{Behavior of \( \tau \) vs. \( \mu \) for fixed values of \( \psi = 0.8, \, \xi = 0.4, \, \rho = 0.15 \). A power law fitting gives a slope of \( \delta = -0.9811(5) \). }
\label{Fig6}
\end{figure}

Figure \ref{Fig6} shows the behavior of \( \tau \) vs. \( \mu \), using fixed values of \( \psi = 0.8, \, \xi = 0.4, \, \rho = 0.15 \). After doing a power law fitting, we find \( \delta = -0.9811(5) \). It is important to mention that a discussion in \cite{teixeira2015convergence} shows analytically that \( \delta = -1 \) is applicable to one-dimensional mappings. This result highlights the significance of the fitting parameter we obtained and its implications for understanding the dynamics of the system near the bifurcation points.

\section{Discussions and conclusions}
\label{sec5}

In our investigation, we did a detailed study of the dynamic properties that control the convergence to the stationary state in a bifurcation with period doubling. Our main focus was on a dissipative version of a kicked rotator system, which is represented by a two-dimensional nonlinear mapping. This mapping provides a way to observe the structure of the system's phase space and the double period bifurcation points once dissipation is included in the system. These bifurcation points show qualitative changes in a dynamical system's behavior as a parameter is varied, indicating transitions between different behaviors such as stability, periodicity, or chaos. Bifurcations in phase space are often marked by the appearance, disappearance, or qualitative change of attractors, trajectories, or other dynamic structures.

Our study specifically looked at the dissipative version of this mapping, which allowed us to examine convergence behaviors both at and near the bifurcation point in detail. By using a scaling law, we could numerically find the critical exponent \( \beta = 1/2 \). This result led us to the values \( \alpha = 1 \) and \( z = -2 \), which agree with expectations from earlier studies in the field.

When we examined the region near the bifurcation more closely, we discovered an exponential decay phenomenon, which indicates a relaxation process. This decay led to the emergence of a relaxation time characterized by the critical exponent \( \delta = -1 \). These findings not only confirm our expectations based on one-dimensional mappings but also align with insights gained from studies on the Fermi-Ulam model \cite{s2023bifurcations,leonel2007scaling,oliveira2009scaling}. This analysis shows the complex nature of dynamical behaviors near bifurcation points and improves our understanding of complex system dynamics.

The study of bifurcations in nonlinear dynamics remains a vibrant and fertile field of research. Future directions may include further investigation of bifurcation phenomena in high-dimensional systems, the development of advanced mathematical techniques for bifurcation analysis \cite{ou2024hopf,tuckerman2000bifurcation}, and the application of bifurcation theory to address real-world problems in science and engineering. By enhancing our understanding of bifurcations, we can explore new horizons in nonlinear dynamics and its diverse applications across various scientific fields.

\section*{Acknowledgements}
D.S.R. expresses gratitude to CAPES, which is a Brazilian agency, and to the University of North Dakota for their hospitality. E.D.L. acknowledges the support received from CNPq (301318/2019-0, 303707/2015-1, and 304398/2023-3) and FAPESP (2021/09519-5 and 2019/14038-6), both of which are Brazilian agencies.

\bibliographystyle{elsarticle-harv} 
\bibliography{example}

\end{document}